\documentclass[aps,prl,superscriptaddress,showpacs,floatfix,twocolumn]{revtex4}

\usepackage{graphicx}   

\begin{document}

\title{Enhanced production of direct photons in Au~+~Au collisions 
at $\sqrt{s_{_{NN}}}$=200~GeV \\
and implications for the initial temperature}

\newcommand{\abilene}{Abilene Christian University, Abilene, TX 79699, USA}
\newcommand{\banaras}{Department of Physics, Banaras Hindu University, Varanasi 221005, India}
\newcommand{\bnl}{Brookhaven National Laboratory, Upton, NY 11973-5000, USA}
\newcommand{\caucr}{University of California - Riverside, Riverside, CA 92521, USA}
\newcommand{\charlesczech}{Charles University, Ovocn\'{y} trh 5, Praha 1, 116 36, Prague, Czech Republic}
\newcommand{\ciae}{China Institute of Atomic Energy (CIAE), Beijing, People's Republic of China}
\newcommand{\cns}{Center for Nuclear Study, Graduate School of Science, University of Tokyo, 7-3-1 Hongo, Bunkyo, Tokyo 113-0033, Japan}
\newcommand{\colorado}{University of Colorado, Boulder, CO 80309, USA}
\newcommand{\columbia}{Columbia University, New York, NY 10027 and Nevis Laboratories, Irvington, NY 10533, USA}
\newcommand{\czechtech}{Czech Technical University, Zikova 4, 166 36 Prague 6, Czech Republic}
\newcommand{\dapnia}{Dapnia, CEA Saclay, F-91191, Gif-sur-Yvette, France}
\newcommand{\debrecen}{Debrecen University, H-4010 Debrecen, Egyetem t{\'e}r 1, Hungary}
\newcommand{\elte}{ELTE, E{\"o}tv{\"o}s Lor{\'a}nd University, H - 1117 Budapest, P{\'a}zm{\'a}ny P. s. 1/A, Hungary}
\newcommand{\fit}{Florida Institute of Technology, Melbourne, FL 32901, USA}
\newcommand{\fsu}{Florida State University, Tallahassee, FL 32306, USA}
\newcommand{\gsu}{Georgia State University, Atlanta, GA 30303, USA}
\newcommand{\hiroshima}{Hiroshima University, Kagamiyama, Higashi-Hiroshima 739-8526, Japan}
\newcommand{\ihepprot}{IHEP Protvino, State Research Center of Russian Federation, Institute for High Energy Physics, Protvino, 142281, Russia}
\newcommand{\illuiuc}{University of Illinois at Urbana-Champaign, Urbana, IL 61801, USA}
\newcommand{\instpasczech}{Institute of Physics, Academy of Sciences of the Czech Republic, Na Slovance 2, 182 21 Prague 8, Czech Republic}
\newcommand{\isu}{Iowa State University, Ames, IA 50011, USA}
\newcommand{\jinrdubna}{Joint Institute for Nuclear Research, 141980 Dubna, Moscow Region, Russia}
\newcommand{\kaeri}{KAERI, Cyclotron Application Laboratory, Seoul, Korea}
\newcommand{\kek}{KEK, High Energy Accelerator Research Organization, Tsukuba, Ibaraki 305-0801, Japan}
\newcommand{\kfki}{KFKI Research Institute for Particle and Nuclear Physics of the Hungarian Academy of Sciences (MTA KFKI RMKI), H-1525 Budapest 114, POBox 49, Budapest, Hungary}
\newcommand{\korea}{Korea University, Seoul, 136-701, Korea}
\newcommand{\kurchatov}{Russian Research Center ``Kurchatov Institute", Moscow, Russia}
\newcommand{\kyoto}{Kyoto University, Kyoto 606-8502, Japan}
\newcommand{\labllr}{Laboratoire Leprince-Ringuet, Ecole Polytechnique, CNRS-IN2P3, Route de Saclay, F-91128, Palaiseau, France}
\newcommand{\lawllnl}{Lawrence Livermore National Laboratory, Livermore, CA 94550, USA}
\newcommand{\losalamos}{Los Alamos National Laboratory, Los Alamos, NM 87545, USA}
\newcommand{\lpc}{LPC, Universit{\'e} Blaise Pascal, CNRS-IN2P3, Clermont-Fd, 63177 Aubiere Cedex, France}
\newcommand{\lund}{Department of Physics, Lund University, Box 118, SE-221 00 Lund, Sweden}
\newcommand{\muenster}{Institut f\"ur Kernphysik, University of Muenster, D-48149 Muenster, Germany}
\newcommand{\myongji}{Myongji University, Yongin, Kyonggido 449-728, Korea}
\newcommand{\nagasaki}{Nagasaki Institute of Applied Science, Nagasaki-shi, Nagasaki 851-0193, Japan}
\newcommand{\newmex}{University of New Mexico, Albuquerque, NM 87131, USA}
\newcommand{\nmsu}{New Mexico State University, Las Cruces, NM 88003, USA}
\newcommand{\ornl}{Oak Ridge National Laboratory, Oak Ridge, TN 37831, USA}
\newcommand{\orsay}{IPN-Orsay, Universite Paris Sud, CNRS-IN2P3, BP1, F-91406, Orsay, France}
\newcommand{\peking}{Peking University, Beijing, People's Republic of China}
\newcommand{\pnpi}{PNPI, Petersburg Nuclear Physics Institute, Gatchina, Leningrad region, 188300, Russia}
\newcommand{\riken}{RIKEN, The Institute of Physical and Chemical Research, Wako, Saitama 351-0198, Japan}
\newcommand{\rikjrbrc}{RIKEN BNL Research Center, Brookhaven National Laboratory, Upton, NY 11973-5000, USA}
\newcommand{\rikkyo}{Physics Department, Rikkyo University, 3-34-1 Nishi-Ikebukuro, Toshima, Tokyo 171-8501, Japan}
\newcommand{\saispbstu}{Saint Petersburg State Polytechnic University, St. Petersburg, Russia}
\newcommand{\saopaulo}{Universidade de S{\~a}o Paulo, Instituto de F\'{\i}sica, Caixa Postal 66318, S{\~a}o Paulo CEP05315-970, Brazil}
\newcommand{\seoulnat}{System Electronics Laboratory, Seoul National University, Seoul, Korea}
\newcommand{\stonybrkc}{Chemistry Department, Stony Brook University, Stony Brook, SUNY, NY 11794-3400, USA}
\newcommand{\stonycrkp}{Department of Physics and Astronomy, Stony Brook University, SUNY, Stony Brook, NY 11794, USA}
\newcommand{\subatech}{SUBATECH (Ecole des Mines de Nantes, CNRS-IN2P3, Universit{\'e} de Nantes) BP 20722 - 44307, Nantes, France}
\newcommand{\tenn}{University of Tennessee, Knoxville, TN 37996, USA}
\newcommand{\titech}{Department of Physics, Tokyo Institute of Technology, Oh-okayama, Meguro, Tokyo 152-8551, Japan}
\newcommand{\tsukuba}{Institute of Physics, University of Tsukuba, Tsukuba, Ibaraki 305, Japan}
\newcommand{\vandy}{Vanderbilt University, Nashville, TN 37235, USA}
\newcommand{\waseda}{Waseda University, Advanced Research Institute for Science and Engineering, 17 Kikui-cho, Shinjuku-ku, Tokyo 162-0044, Japan}
\newcommand{\weizmann}{Weizmann Institute, Rehovot 76100, Israel}
\newcommand{\yonsei}{Yonsei University, IPAP, Seoul 120-749, Korea}
\affiliation{\abilene}
\affiliation{\banaras}
\affiliation{\bnl}
\affiliation{\caucr}
\affiliation{\charlesczech}
\affiliation{\ciae}
\affiliation{\cns}
\affiliation{\colorado}
\affiliation{\columbia}
\affiliation{\czechtech}
\affiliation{\dapnia}
\affiliation{\debrecen}
\affiliation{\elte}
\affiliation{\fit}
\affiliation{\fsu}
\affiliation{\gsu}
\affiliation{\hiroshima}
\affiliation{\ihepprot}
\affiliation{\illuiuc}
\affiliation{\instpasczech}
\affiliation{\isu}
\affiliation{\jinrdubna}
\affiliation{\kaeri}
\affiliation{\kek}
\affiliation{\kfki}
\affiliation{\korea}
\affiliation{\kurchatov}
\affiliation{\kyoto}
\affiliation{\labllr}
\affiliation{\lawllnl}
\affiliation{\losalamos}
\affiliation{\lpc}
\affiliation{\lund}
\affiliation{\muenster}
\affiliation{\myongji}
\affiliation{\nagasaki}
\affiliation{\newmex}
\affiliation{\nmsu}
\affiliation{\ornl}
\affiliation{\orsay}
\affiliation{\peking}
\affiliation{\pnpi}
\affiliation{\riken}
\affiliation{\rikjrbrc}
\affiliation{\rikkyo}
\affiliation{\saispbstu}
\affiliation{\saopaulo}
\affiliation{\seoulnat}
\affiliation{\stonybrkc}
\affiliation{\stonycrkp}
\affiliation{\subatech}
\affiliation{\tenn}
\affiliation{\titech}
\affiliation{\tsukuba}
\affiliation{\vandy}
\affiliation{\waseda}
\affiliation{\weizmann}
\affiliation{\yonsei}
\author{A.~Adare}	\affiliation{\colorado}
\author{S.~Afanasiev}	\affiliation{\jinrdubna}
\author{C.~Aidala}	\affiliation{\columbia}
\author{N.N.~Ajitanand}	\affiliation{\stonybrkc}
\author{Y.~Akiba}	\affiliation{\riken} \affiliation{\rikjrbrc}
\author{H.~Al-Bataineh}	\affiliation{\nmsu}
\author{J.~Alexander}	\affiliation{\stonybrkc}
\author{A.~Al-Jamel}	\affiliation{\nmsu}
\author{K.~Aoki}	\affiliation{\kyoto} \affiliation{\riken}
\author{L.~Aphecetche}	\affiliation{\subatech}
% ...
\collaboration{PHENIX Collaboration} \noaffiliation

\date{\today}

\begin{abstract}

The production of $e^+e^-$ pairs for $m_{e^+e^-}<0.3$~GeV/$c^2$ 
and $1<p_T<5$~GeV/$c$ is measured in $p+p$ and Au~+~Au 
collisions at $\sqrt{s_{_{NN}}}=200$~GeV.  Enhanced yield above 
hadronic sources is observed.  Treating the excess as photon 
internal conversions, the invariant yield of direct photons is 
deduced.  In central Au~+~Au collisions, the excess of direct 
photon yield over $p+p$ is exponential in transverse momentum, 
with inverse slope $T=221 \pm 19^{\rm stat} \pm 19^{\rm 
syst}$~MeV.  Hydrodynamical models with initial temperatures 
ranging from $T_{\rm init}\sim$300--600 MeV at times of 
$\sim$0.6--0.15 fm/$c$ after the collision are in qualitative 
agreement with the data.  Lattice QCD predicts a phase transition 
to quark gluon plasma at $\sim$170~MeV.

\end{abstract}


\maketitle


Experimental results from the Relativistic Heavy Ion Collider (RHIC) have 
established the formation of dense partonic matter in Au~+~Au collisions 
at $\sqrt{s_{NN}}=200$ GeV~\cite{Adcox:2004mh}.  The large energy loss of 
light quarks and gluons~\cite{Adcox:2001jp} as well as that of heavy 
quarks~\cite{Adare:2006nq} indicates that the matter is very dense.  The 
strong elliptic flow of light~\cite{Ackermann:2000tr,Adler:2003kt} and 
charmed~\cite{Adare:2006nq} hadrons indicates rapid thermalization.  Such a 
hot, dense medium should emit thermal radiation~\cite{Stankus:2005eq}; the 
partonic phase is predicted to be the dominant source of direct photons 
with $1<p_T<3$ GeV/$c$ in Au~+~Au collisions at 
RHIC~\cite{Turbide:2003si}.

Observation of thermal photons will allow determination of the initial 
temperature of the matter.  However, the measurement of direct photons for 
$1<p_T<3$ GeV/$c$ is notoriously difficult due to a large background from 
hadronic decay photons.  Direct photons contribute only $\simeq$ 10\% above 
the background photon yield~\cite{Turbide:2003si}.  In general, any source 
of high energy photons can also emit virtual photons, which convert to low 
mass $e^+e^-$ pairs.  For example, gluon Compton scattering ($q+g 
\rightarrow q+\gamma$) has an associated process that produces low mass 
$e^+e^-$ pairs through internal conversion ($q+g \rightarrow q+\gamma^{*} 
\rightarrow q+e^+e^-$).  Consequently, we search for ``quasi-real'' virtual 
photons, which appear as low invariant mass $e^+e^-$ pairs.

The relation between photon production 
and the associated $e^+e^-$ pair production 
can be written as \cite{Lichard:1994yx,ppg088}
\begin{eqnarray}
\frac{d^2n_{ee}}{dm_{ee}} = \frac{2\alpha}{3\pi}\frac{1}{m_{ee}} \sqrt{1-\frac{4m_e^2}{m_{ee}^2}}\Bigl( 1+\frac{2m_e^2}{m_{ee}^2} \Bigr) S  dn_{\gamma} \label{eq:Conversion}
\end{eqnarray}
Here $\alpha$ is the fine structure constant,
$m_e$ and $m_{ee}$ are the masses of the electron and the $e^+e^-$ pair respectively,
and $S$ is a process dependent factor that goes to 1 as
$m_{ee} \rightarrow 0$ or $m_{ee} \ll p_T$.
Equation~(\ref{eq:Conversion}) also describes the relation between
the photons from hadron decays
(e.g.  $\pi^0, \eta \rightarrow \gamma \gamma$, and
$\omega \rightarrow \gamma \pi^0$)
and the $e^+e^-$ pairs from Dalitz decays (
$\pi^0, \eta \rightarrow e^+e^-\gamma$ and $\omega \rightarrow e^+e^-\pi^0$).
For $\pi^0$ and $\eta$, the factor $S$ is given by
$S =|F(m_{ee}^2)|^2 (1-\frac{m_{ee}^2}{M_h^2})^3$~\cite{Landsberg:1986fd}, 
where $M_h$ is the meson mass and $F(m_{ee}^2)$ is the 
form factor.

The factor $S$ for a hadron $h$ is zero for $m_{ee} > M_h$.
We exploit this cut-off to separate the direct photon signal
from the hadronic background.  Since 80\% of the hadronic photons are
from $\pi^0$ decays,
the signal to background (S/B) ratio for the direct photon 
signal improves by a factor of five
for $m_{ee}> M_{\pi^0}$=135~MeV/$c^2$, thereby allowing a direct photon
signal that is 10\% of the 
background to be observed as a 50\% excess of $e^+e^-$ pairs.

In this Letter we present the analysis of $e^+e^-$ pairs
for $m_{ee} < 0.3$~GeV/$c^2$ and for $1<p_T<5$~GeV/$c$
in Au~+~Au and $p+p$ collisions at $\sqrt{s_{_{NN}}}=200$ GeV
recorded during 2004 and 2005, respectively.
The PHENIX detector~\cite{Adcox:2003zm} measures electrons in the two central
arms, each covering $\left|\Delta\eta\right|\le0.35$ in pseudorapidity
and $\pi/2$ in azimuthal angle.  
The Au~+~Au analysis~\cite{ppg088,ppg075}
uses 8$\times 10^8$ minimum bias (Min. Bias) events
corresponding to $92.2^{+2.5}_{-3.0}$\% of the
inelastic Au~+~Au cross section.
The beam-beam counters and zero degree calorimeters provide
the Min. Bias trigger, as well as the centrality
selection\cite{Adler:2003cb}.
The $p+p$ analysis~\cite{ppg085} uses
43 nb$^{-1}$ of
data recorded using the Min. Bias trigger
and 2.25 pb$^{-1}$ of single electron triggered data.
Helium bags in both runs 
reduced the total conversion material, including the beam pipe,
to $\sim$0.4\% of a radiation length.

All electrons and positrons with $p_T >$ 0.2 GeV/$c$ are combined into
pairs.  
Pairs from photon conversions in the detector material are
removed by a cut on the orientation of the pair in the magnetic
field~\cite{ppg088}.
The combinatorial background is computed 
by mixing events and is subtracted~\cite{ppg088,ppg075}.
The S/B ratio is $\sim$0.2~(at $m_{ee}$=0.3~GeV) 
to $\sim$1.5~ (at $m_{ee}=0.1~{\rm GeV}/c^2$) 
for $p_T>2$~GeV/$c$ and ~0.05 to 0.4 for $1< p_T <2$~GeV/$c$.
There are two sources of correlated background: 
two $e^+e^-$ pairs from a meson decay and correlated hadrons
decaying into two $e^+e^-$ pairs, either within the same jet or in
back-to-back jets.  The 
magnitude of the correlated background,
about 10\% of the signal in $p+p$, is determined from the
like-sign pair data and subtracted
after correcting for acceptance differences between like and
unlike-sign pairs~\cite{ppg085}.
We correct for electron reconstruction
efficiency, and in $p+p$ for trigger efficiency, determined as
a function of mass and pair $p_T$ using a GEANT-based
Monte Carlo simulation~\cite{GEANT} of the PHENIX detector.

\begin{figure}[tb]
\includegraphics[width=1.0\linewidth]{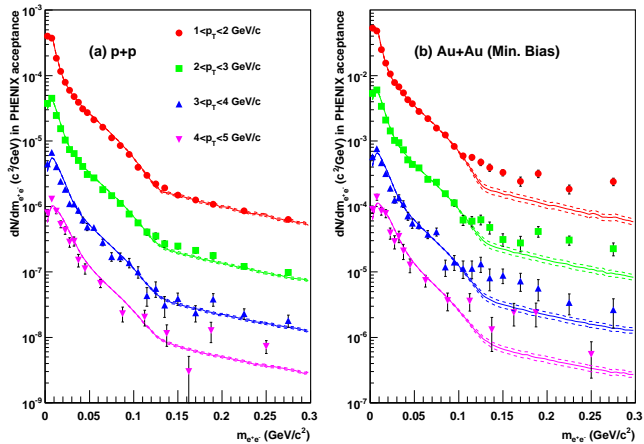}
\caption{\label{fig:mass} (color online)
The measured $e^+e^-$ pair invariant mass distributions.
The $p_T$ ranges are shown in the legend.
The solid curves represent an estimate of hadronic sources; the
dashed curves represent the uncertainty in the estimate.
}
\end{figure}

Figure~\ref{fig:mass} shows the 
mass spectra of
$e^+e^-$ pairs in $p+p$ and Au~+~Au collisions for different ranges of
pair $p_T$, comparing to 
a ``cocktail'' of hadron decays calculated using
a Monte Carlo hadron decay generator based on meson production measured by PHENIX~\cite{ppg088}.
Detector resolution 
is included in the cocktail calculation.
The open charm contribution, calculated with
{\sc PYTHIA}~\cite{Sjostrand:2001}, is also included 
but is negligible in this kinematic range.
The cocktail is normalized to the data for $m_{ee}<0.03$~GeV/$c^2$;
the absolute normalization agrees with the data within a
20\% systematic uncertainty~\cite{ppg075,ppg085}.
The ``knee'' beginning at $m_{ee} \simeq 0.1$~GeV/$c^2$ corresponds to the
$\pi^0$ cut-off, leading to an 80\% reduction of background
above this point.
The $p+p$ data are consistent with the background
for $m_{ee} \geq M_{\pi^0}$ at 
lower $p_T$, but reveal 
a small excess over the background at higher $p_T$.
A much greater excess is observed in Au~+~Au 
indicating enhanced production of
virtual photons.

Internal conversion of direct photons
is a possible source of the excess.
Little contribution from other sources
of $e^+e^-$ pairs is expected in this 
mass region since
$\pi^{+}\pi^{-} \rightarrow e^+e^-$
can only contribute for $m_{ee} \ge 2 M_{\pi}$.
Although PHENIX has observed a strong enhancement of $e^+e^-$ 
pairs for $0.15<m_{ee}<0.75$~GeV/$c^2$ in Au~+~Au, it peaks at 
low $p_T$ and decreases rapidly with increasing $p_T$\cite{ppg088}
with a different mass distribution than that observed at high 
$p_T$.  

\begin{figure}[tb]
\includegraphics[width=1.0\linewidth]{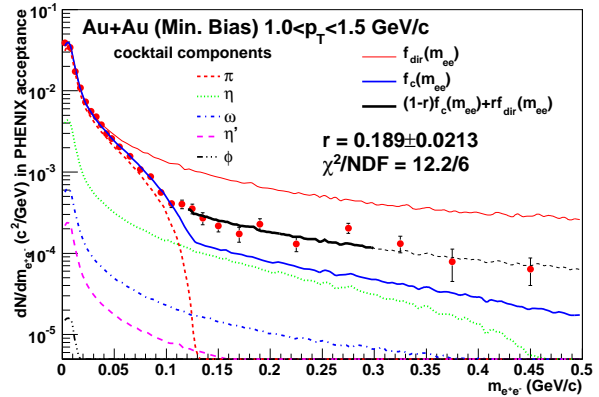}
\caption{\label{fig:fit} (color online)
Electron pair mass distribution for Au~+~Au (Min. Bias) events
for $1.0<p_T<1.5$ GeV/$c$.  The two-component fit is explained
in the text.  The fit range is $0.12<m_{ee}<0.3$~GeV/$c^2$.  
The dashed (black) curve at greater $m_{ee}$ shows 
$f(m_{ee})$ outside of the fit range.
}
\end{figure}

Figure~\ref{fig:fit} shows that the mass spectrum for 
$m_{ee}<0.5$~GeV/$c^2$ and $p_T>1$~GeV/$c$ is well described by 
the cocktail plus internal conversion photons.  The flat mass 
spectrum of the excess above the cocktail at this $p_T$ shows no 
significant indication of low-mass enhancement~\cite{ppg088}.
Thus, we treat
the excess entirely as internal conversion of
photons and deduce the real direct photon yield from
$e^+e^-$ pairs using Eq.~(\ref{eq:Conversion}).

We fit a two-component
function 
$f(m_{ee})=(1-r)f_{c}(m_{ee}) + r~f_{\rm dir}(m_{ee})$,
to the mass distribution.
$f_{c}(m_{ee})$ is the shape of the cocktail
mass distribution (shown in Fig.~\ref{fig:mass}),
$f_{\rm dir}(m_{ee})$ is the expected
shape of the direct photon internal conversion, and
$r$ is the fit parameter.
We assume that the form factor for direct photons
is $F(m_{ee}^2)=1$, as one would expect from a purely point-like process.
For direct photons from parton fragmentation or from hadronic
gas, $F(m_{ee}^2)$ may be greater than one.
If we arbitrarily set the form factor in $f_{\rm dir}(m_{ee})$ to be
the same as that in $f_{\eta}(m_{ee})$, $r$ would decrease by $\simeq$ 10\%.

For each $p_T$ bin, $f(m_{ee})$ is fit to the data for
$m_{\rm low} < m_{ee} < 0.3$~GeV/$c^2$ with $m_{\rm low}=0.08,0.1,0.12$~GeV/$c^2$;
$r$ is the only fit parameter.
Figure~\ref{fig:fit} shows $f_{\rm dir}(m_{ee})$ and 
$f_{c}(m_{ee})$ together with
a fit result for Au~+~Au (Min. Bias) data for $1.0<p_T<1.5$ GeV/$c$.
For higher $p_T$ bins, $\chi^2/NDF$ is near 1.0; fits to
centrality separated data also give good $\chi^2/NDF$.

Therefore, we
focus on the uncertainties that can cause 
distortions in the mass distribution, namely (i) the
particle composition in the hadronic background,
(ii) the background (from mixed events and correlated pairs), 
(iii) the geometric acceptance due to detector active areas, and
(iv) the efficiency corrections.
These were studied by Monte Carlo simulation.  
The mass spectrum
is distorted within the systematic uncertainties,
and the fitting procedure is applied to the distorted
spectrum to determine the systematic uncertainties in $r$.
The systematic uncertainty due to the variation of $m_{\rm low}$
is also included.
The dominant uncertainty is the particle composition
in the hadronic cocktail, namely the $\eta/\pi^0$ ratio which is 
$0.48 \pm 0.03 (0.08)$ at high $p_T$ for $p+p$ (Au~+~Au) based on 
PHENIX measurements~\cite{Adler:2006hu}.
This corresponds to a $\simeq 7$\% ($\simeq 17$\%) uncertainty in
the $p+p$ (Au~+~Au) cocktail for $0.1 < m_{ee} < 0.3$~GeV/$c^2$.  Other 
sources cause
only a few percent uncertainty in the data to cocktail ratio.  

\begin{figure}[tb]
\includegraphics[width=1.0\linewidth]{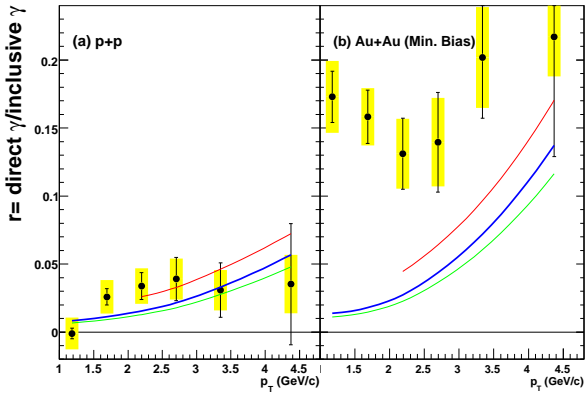}
\caption{\label{fig:ratio} (color online)
The fraction of the direct photon component as
a function of $p_T$.  
The error bars and the error band represent the statistical and
systematic uncertainties, respectively.  The curves are
from a NLO pQCD calculation (see text).
}
\end{figure}

Figure~\ref{fig:ratio} shows
the fraction $r$ of the direct photon component determined by the 
two-component fit in (a) $p+p$ and (b) Au~+~Au (Min. Bias).
The curves represent the expectations from a next-to-leading-order
perturbative QCD (NLO pQCD) calculation~\cite{Gordon:1993qc}.
For $p+p$, the curves show the ratio
$d\sigma^{NLO}_{\gamma}(p_T)/d\sigma^{\rm incl}_{\gamma}(p_T)$,
where $d\sigma^{NLO}_{\gamma}(p_T)$ is the direct photon
cross section from the NLO pQCD calculation and
$d\sigma^{\rm incl}_{\gamma}(p_T)$ is the inclusive photon cross section.
For Au~+~Au, the curves represent
$T_{\rm AA}d\sigma^{NLO}_{\gamma}(p_T)/dN^{\rm incl}_{\gamma}(p_T)$,
where $T_{\rm AA}$
is the Glauber nuclear overlap
function 
and $dN^{\rm incl}_{\gamma}(p_T)$ is the inclusive photon yield.
The three curves correspond, from top to bottom,
to the theory scale  $\mu$ = 0.5~$p_T$, $p_T$, and 2~$p_T$, 
respectively, showing the scale dependence of the theory.
While the fraction $r$ is consistent 
with the NLO pQCD calculation~\cite{Gordon:1993qc} in $p+p$, it is
larger than the calculation in Au~+~Au for $p_T <3.5$ GeV/$c$.

\begin{figure}[tb]
\includegraphics[width=1.0\linewidth]{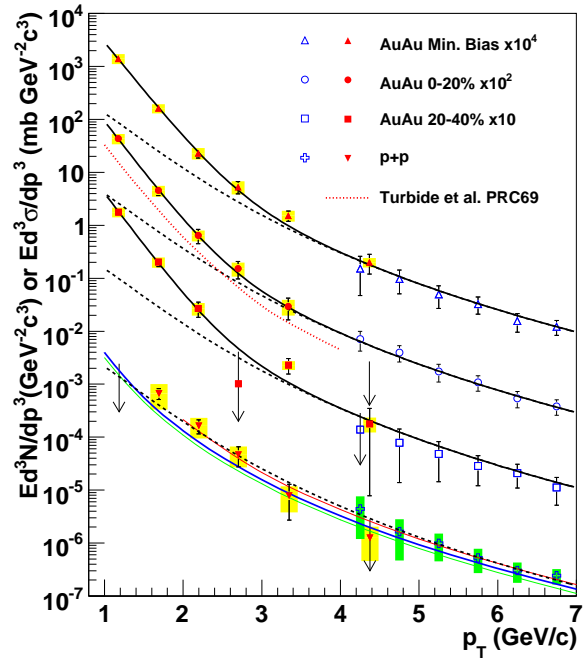}
\caption{\label{fig:spectra} (color online)
Invariant cross section ($p+p$) and invariant yield (Au~+~Au) of
direct photons as a function of $p_T$.  The filled points are from
this analysis and open points are from~\cite{Adler:2005ig,Adler:2006yt}.
The three curves on the $p+p$ data represent NLO pQCD 
calculations, and the dashed curves show a modified power-law 
fit to the $p+p$ data, scaled by $T_{\rm AA}$.  The dashed (black) 
curves are exponential plus the $T_{\rm AA}$ scaled 
$p+p$ fit.   The dotted (red) curve near the 0--20\% centrality 
data is a theory calculation~\cite{Turbide:2003si}.
}
\end{figure}

The direct photon fraction $r$ in Fig.~\ref{fig:ratio} is converted
to the direct photon yield as
$dN^{\rm dir}(p_T)= r \times dN^{\rm incl}(p_T)$.
The inclusive photon yield $dN^{\rm incl}(p_T)$ for each
$p_T$ bin is determined 
from the yield of $e^+e^-$ pairs for $m_{ee}<0.03$~GeV/$c^2$
using Eq.~(\ref{eq:Conversion}).
Here we use the fact that in this mass range the process dependent
factor $S$ is unity within a few percent for
any photon source.

Figure~\ref{fig:spectra} compares the direct photon spectra with 
previously measured direct photon data 
from~\cite{Adler:2005ig,Adler:2006yt} and 
NLO pQCD calculations~\cite{Gordon:1993qc}.
The systematic uncertainty of the inclusive photon (14\%
from the uncertainty in the $e^+e^-$ pair acceptance correction\cite{ppg075}) is
added in quadrature with the systematic uncertainties of these data.
The $p+p$ data are shown as an invariant cross section
using $d\sigma = \sigma^{\rm inel}_{pp} dN$.

In this analysis we have converted the yield of excess
$e^+e^-$ pairs to that of real direct photons using 
Eq.~(\ref{eq:Conversion}), assuming $S=1$.
This implies $\frac{d^2n_{ee}}{dm_{ee}} = 
\frac{2\alpha}{3\pi}\frac{1}{m_{ee}} dn_{\gamma}$.  Thus
the yield of the excess $e^+e^-$ pairs for 
$0.1<m_{ee}<0.3$~GeV$/c^2$ before the conversion can be obtained by 
multiplying the direct photon yield by a factor of
$\frac{2\alpha}{3\pi} \log{\frac{300}{100}} = 1.7 \times 10^{-3}$.

The pQCD calculation is consistent with the $p+p$ data within
the theoretical uncertainties for $p_T>2$ GeV/$c$.
A similarly good agreement is observed
for $\pi^0$~\cite{Adler:2003pb}.
The $p+p$ data can be well described by a modified power-law
function ($A_{pp} (1+p_T^2/b)^{-n}$) as shown by the dashed curve
in Fig.~\ref{fig:spectra}.  
The Au~+~Au data are
above the~$p+p$~fit curve scaled by $T_{\rm AA}$ for
$p_T<2.5$ GeV/$c$, indicating that the direct photon yield in the low
$p_T$ range increases faster than the 
binary $NN$ collision scaled $p+p$ cross section.

\begin{table}[thb]
\caption{Summary of the fits.  The first and second errors are
statistical and systematic, respectively.}
\begin{ruledtabular}
\begin{tabular}{lccc}
centrality & $dN/dy$($p_T>1$GeV/$c$) &$T$(MeV)     &$\chi^2/DOF$\\
\hline
 0-20\%& $1.50\pm0.23\pm0.35$ & $221\pm19\pm19$ & 4.7/4\\
20-40\%& $0.65\pm0.08\pm0.15$ & $217\pm18\pm16$  & 5.0/3\\
Min. Bias     & $0.49\pm0.05\pm0.11$ & $233\pm14\pm19$ & 3.2/4\\
\end{tabular}
\end{ruledtabular}
\label{tab:summary}
\end{table}

We fit an exponential plus the $T_{\rm AA}$-scaled $p+p$ fit function
($A e^{-p_T/T} + T_{\rm AA} \times A_{pp} (1+p_T^2/b)^{-n}$) to the Au~+~Au data.
The only free parameters in the fit are 
$A$ and the inverse slope $T$ of the exponential term.
The systematic uncertainties in $T$ are estimated
by changing the $p+p$ fit component and the Au~+~Au data points
within the systematic uncertainties.
The results of the fits are summarized in Table~\ref{tab:summary},
where $A$ is converted to $dN/dy$ for $p_T>1$GeV/$c$.
For central collisions $T=221~\pm~19^{\rm stat}~\pm~19^{\rm syst}$~MeV.
Using, instead, a power-law function ($\propto p_T^{-n}$) 
to fit the $p+p$ spectrum
yields $n = 5.40 \pm 0.15$, and 
$T_{\rm AuAu}=240 \pm 21$ MeV.
If the direct photons in Au~+~Au collisions are of thermal origin,
the inverse slope $T$ is related to the initial temperature
$T_{\rm init}$ of the dense matter.  In hydrodynamical models,
$T_{\rm init}$ is 1.5 to 3 times $T$ due to the space-time
evolution~\cite{d'Enterria:2005vz}.
Several hydrodynamical models can reproduce the central Au~+~Au data within 
a factor of two~\cite{ppg088}.
These assume formation of a hot system with initial temperature ranging
from $T_{\rm init} =  300$ MeV at thermalization time $\tau_0 = 0.6$ fm/$c$
to $T_{\rm init} = 600$ MeV at $\tau_0=0.15$ fm/$c$~\cite{d'Enterria:2005vz}.
As an example, the dotted (red) curve in Fig.~\ref{fig:spectra} 
shows a thermal photon spectrum in central Au~+~Au collisions
calculated with $T_{\rm init}=370$ MeV~\cite{Turbide:2003si}.

In conclusion, we have measured $e^+e^-$ pairs with
$m_{ee}<300$ MeV/$c^2$ and $1<p_T<5$ GeV/$c$ in $p+p$ and Au~+~Au
collisions.  The $p+p$ data show 
a small excess over
the hadronic background while the Au~+~Au data show a much larger excess.
By treating the excess as internal conversion of
direct photons, the direct photon yield is deduced.
The yield is consistent with a NLO pQCD
calculation in $p+p$.
In central Au~+~Au collisions
the shape of the direct photon spectrum above the 
$T_{\rm AA}$-scaled~$p+p$~spectrum 
is exponential in $p_T$, with an 
inverse slope 
$T=221 \pm 19^{\rm stat} \pm 19^{\rm syst}$~MeV.
Hydrodynamical models with $T_{\rm init}\sim$300--600 MeV
at $\tau_0 \sim$0.6--0.15 fm/$c$
are in qualitative agreement with the data.
Lattice QCD predicts a phase transition from hadronic phase to
quark gluon plasma
at $\sim$170 MeV\cite{Adcox:2004mh}.


We thank the staff of the Collider-Accelerator and 
Physics Departments at BNL for their vital contributions.
We acknowledge support from the Department of Energy and NSF (USA),
MEXT and JSPS (Japan), CNPq and FAPESP (Brazil), NSFC (China),
MSMT (Czech Republic), IN2P3/CNRS, and CEA (France),
BMBF, DAAD, and AvH (Germany), OTKA (Hungary), DAE (India),
ISF (Israel), NRF (Korea), MES, RAS, and FAAE (Russia),
VR and KAW (Sweden), U.S. CRDF for the FSU, US-Hungarian
NSF-OTKA-MTA, and US-Israel BSF.

\def\IJMPA{{Int. J. Mod. Phys.}~{\bf A}}
\def\JPG{{J. Phys}~{\bf G}}
\def\NCA{Nuovo Cimento}
\def\NIM{Nucl. Instrum. Methods}
\def\NIMA{{Nucl. Instrum. Methods}~{\bf A}}
\def\NPA{{Nucl. Phys.}~{\bf A}}
\def\NPB{{Nucl. Phys.}~{\bf B}}
\def\PLB{Phys. Lett. B}
\def\PLC{Phys. Repts.\ }
\def\PRL{Phys. Rev. Lett.\ }
\def\PRD{Phys. Rev. D}
\def\PRC{Phys. Rev. C}
\def\ZPC{{Z. Phys.}~{\bf C}}
\def\etal{{\it et al.}}


\begin{thebibliography}{22}
\expandafter\ifx\csname natexlab\endcsname\relax\def\natexlab#1{#1}\fi
\expandafter\ifx\csname bibnamefont\endcsname\relax
  \def\bibnamefont#1{#1}\fi
\expandafter\ifx\csname bibfnamefont\endcsname\relax
  \def\bibfnamefont#1{#1}\fi
\expandafter\ifx\csname citenamefont\endcsname\relax
  \def\citenamefont#1{#1}\fi
\expandafter\ifx\csname url\endcsname\relax
  \def\url#1{\texttt{#1}}\fi
\expandafter\ifx\csname urlprefix\endcsname\relax\def\urlprefix{URL }\fi
\providecommand{\bibinfo}[2]{#2}
\providecommand{\eprint}[2][]{\url{#2}}

\bibitem[{\citenamefont{Adcox et~al.}(2005)}]{Adcox:2004mh}
\bibinfo{author}{\bibfnamefont{K.}~\bibnamefont{Adcox}} \bibnamefont{et~al.},
  \bibinfo{journal}{Nucl. Phys. A} \textbf{\bibinfo{volume}{757}},
  \bibinfo{pages}{184} (\bibinfo{year}{2005}).

\bibitem[{\citenamefont{Adcox et~al.}(2002)}]{Adcox:2001jp}
\bibinfo{author}{\bibfnamefont{K.}~\bibnamefont{Adcox}} \bibnamefont{et~al.},
  \bibinfo{journal}{Phys. Rev. Lett.} \textbf{\bibinfo{volume}{88}},
  \bibinfo{pages}{022301} (\bibinfo{year}{2002}).

\bibitem[{\citenamefont{Adare et~al.}(2007)}]{Adare:2006nq}
\bibinfo{author}{\bibfnamefont{A.}~\bibnamefont{Adare}} \bibnamefont{et~al.},
  \bibinfo{journal}{Phys. Rev. Lett.} \textbf{\bibinfo{volume}{98}},
  \bibinfo{pages}{172301} (\bibinfo{year}{2007}).

\bibitem[{\citenamefont{Ackermann et~al.}(2001)}]{Ackermann:2000tr}
\bibinfo{author}{\bibfnamefont{K.~H.} \bibnamefont{Ackermann}}
  \bibnamefont{et~al.}, \bibinfo{journal}{Phys. Rev. Lett.}
  \textbf{\bibinfo{volume}{86}}, \bibinfo{pages}{402} (\bibinfo{year}{2001}).

\bibitem[{\citenamefont{Adler et~al.}(2003{\natexlab{a}})}]{Adler:2003kt}
\bibinfo{author}{\bibfnamefont{S.~S.} \bibnamefont{Adler}}
  \bibnamefont{et~al.}, \bibinfo{journal}{Phys. Rev. Lett.}
  \textbf{\bibinfo{volume}{91}}, \bibinfo{pages}{182301}
  (\bibinfo{year}{2003}{\natexlab{a}}).

\bibitem[{\citenamefont{Stankus}(2005)}]{Stankus:2005eq}
\bibinfo{author}{\bibfnamefont{P.}~\bibnamefont{Stankus}},
  \bibinfo{journal}{Ann. Rev. Nucl. Part. Sci.} \textbf{\bibinfo{volume}{55}},
  \bibinfo{pages}{517} (\bibinfo{year}{2005}).

\bibitem[{\citenamefont{Turbide et~al.}(2004)}]{Turbide:2003si}
\bibinfo{author}{\bibfnamefont{S.}~\bibnamefont{Turbide}}, 
\bibinfo{author}{\bibfnamefont{R.}~\bibnamefont{Rapp}},
\bibnamefont{and}
\bibinfo{author}{\bibfnamefont{C.}~\bibnamefont{Gale}},
  \bibinfo{journal}{Phys. Rev.} \textbf{\bibinfo{volume}{C69}},
  \bibinfo{pages}{014903} (\bibinfo{year}{2004}).

\bibitem[{\citenamefont{Lichard}(1995)}]{Lichard:1994yx}
\bibinfo{author}{\bibfnamefont{P.}~\bibnamefont{Lichard}},
  \bibinfo{journal}{Phys. Rev.} \textbf{\bibinfo{volume}{D51}},
  \bibinfo{pages}{6017} (\bibinfo{year}{1995}).

\bibitem[{ppg()}]{ppg088}
\bibinfo{note}{A.~Adare et al.,}.
\eprint{arXiv:0911.0244 [nucl-ex]}.

\bibitem[{\citenamefont{Landsberg}(1985)}]{Landsberg:1986fd}
\bibinfo{author}{\bibfnamefont{L.~G.} \bibnamefont{Landsberg}},
  \bibinfo{journal}{Phys. Rept.} \textbf{\bibinfo{volume}{128}},
  \bibinfo{pages}{301} (\bibinfo{year}{1985}).

\bibitem[{\citenamefont{Adcox et~al.}(2003)}]{Adcox:2003zm}
\bibinfo{author}{\bibfnamefont{K.}~\bibnamefont{Adcox}} \bibnamefont{et~al.},
  \bibinfo{journal}{Nucl. Instrum. Meth.} \textbf{\bibinfo{volume}{A499}},
  \bibinfo{pages}{469} (\bibinfo{year}{2003}).

\bibitem[{\citenamefont{Afanasiev et~al.}(2007)}]{ppg075}
\bibinfo{author}{\bibfnamefont{S.}~\bibnamefont{Afanasiev}}
\bibnamefont{et~al.} 
\eprint{arXiv:0706.3034 [nucl-ex]}.

\bibitem[{\citenamefont{Adler et~al.}(2004)}]{Adler:2003cb}
\bibinfo{author}{\bibfnamefont{S.~S.} \bibnamefont{Adler}}
  \bibnamefont{et~al.}, \bibinfo{journal}{Phys. Rev.}
  \textbf{\bibinfo{volume}{C69}}, \bibinfo{pages}{034909}
  (\bibinfo{year}{2004}).

\bibitem[{\citenamefont{Adare et~al.}(2009)}]{ppg085}
\bibinfo{author}{\bibfnamefont{A.}~\bibnamefont{Adare}} \bibnamefont{et~al.},
  \bibinfo{journal}{Phys. Lett.} \textbf{\bibinfo{volume}{B670}},
  \bibinfo{pages}{313} (\bibinfo{year}{2009}).

\bibitem[{GEA()}]{GEANT}
\bibinfo{note}{{GEANT3.21 CERN Program Library}}.

\bibitem[{\citenamefont{{Sj\"{o}strand} et~al.}(2001)}]{Sjostrand:2001}
\bibinfo{author}{\bibfnamefont{T.}~\bibnamefont{{Sj\"{o}strand}}}
  \bibnamefont{et~al.}, \bibinfo{journal}{Comp. Phys. Comm.}
  \textbf{\bibinfo{volume}{135}}, \bibinfo{pages}{238} (\bibinfo{year}{2001}).

\bibitem[{\citenamefont{Adler et~al.}(2006)}]{Adler:2006hu}
\bibinfo{author}{\bibfnamefont{S.~S.} \bibnamefont{Adler}}
  \bibnamefont{et~al.}, \bibinfo{journal}{Phys. Rev. Lett.}
  \textbf{\bibinfo{volume}{96}}, \bibinfo{pages}{202301}
  (\bibinfo{year}{2006}).

\bibitem[{\citenamefont{Gordon and Vogelsang}(1993)}]{Gordon:1993qc}
\bibinfo{author}{\bibfnamefont{L.~E.} \bibnamefont{Gordon}} \bibnamefont{and}
  \bibinfo{author}{\bibfnamefont{W.}~\bibnamefont{Vogelsang}},
  \bibinfo{journal}{Phys. Rev.} \textbf{\bibinfo{volume}{D48}},
  \bibinfo{pages}{3136} (\bibinfo{year}{1993});
\bibinfo{note}{W.~Vogelsang calculated the cross section.}

\bibitem[{\citenamefont{Adler et~al.}(2005)}]{Adler:2005ig}
\bibinfo{author}{\bibfnamefont{S.~S.} \bibnamefont{Adler}}
  \bibnamefont{et~al.}, \bibinfo{journal}{Phys. Rev. Lett.}
  \textbf{\bibinfo{volume}{94}}, \bibinfo{pages}{232301}
  (\bibinfo{year}{2005}).

\bibitem[{\citenamefont{Adler et~al.}(2007)}]{Adler:2006yt}
\bibinfo{author}{\bibfnamefont{S.~S.} \bibnamefont{Adler}}
  \bibnamefont{et~al.}, \bibinfo{journal}{Phys. Rev. Lett.}
  \textbf{\bibinfo{volume}{98}}, \bibinfo{pages}{012002}
  (\bibinfo{year}{2007}).

\bibitem[{\citenamefont{Adler et~al.}(2003{\natexlab{b}})}]{Adler:2003pb}
\bibinfo{author}{\bibfnamefont{S.~S.} \bibnamefont{Adler}}
  \bibnamefont{et~al.}, \bibinfo{journal}{Phys. Rev. Lett.}
  \textbf{\bibinfo{volume}{91}}, \bibinfo{pages}{241803}
  (\bibinfo{year}{2003}{\natexlab{b}}).

\bibitem[{\citenamefont{d'Enterria and Peressounko}(2006)}]{d'Enterria:2005vz}
\bibinfo{author}{\bibfnamefont{D.}~\bibnamefont{d'Enterria}} \bibnamefont{and}
  \bibinfo{author}{\bibfnamefont{D.}~\bibnamefont{Peressounko}},
  \bibinfo{journal}{Eur. Phys. J.} \textbf{\bibinfo{volume}{C46}},
  \bibinfo{pages}{451} (\bibinfo{year}{2006}) 
  \bibinfo{note}{and references therein.}

\end{thebibliography}

\end{document}